\documentclass[%
reprint,amsmath,amssymb,aps,prl,superscriptaddress,longbibliography,
floatfix,
nobalancelastpage
]{revtex4-2}

\usepackage{graphicx}
\usepackage{dcolumn}
\usepackage{bm}
\usepackage{epstopdf}
\usepackage{float}
\usepackage{braket}
\usepackage{dsfont}
\usepackage{amsmath}
\usepackage{siunitx} 
\usepackage[italicdiff]{physics}
\usepackage[T1]{fontenc}
\usepackage[dvipsnames]{xcolor}
\usepackage{color,soul}
\usepackage{lipsum}
\usepackage[colorlinks=true, allcolors=blue]{hyperref}

\usepackage{float}

\graphicspath{{figures/}}

\usepackage{gensymb}
\usepackage{textcomp}

\begin{document}

\title{Pyroelectric effects in hybrid semiconductor-lithium niobate quantum devices}

\author{Manas Ranjan Sahu}
\thanks{These authors contributed equally.}

\author{Suraj Thapa Magar}
\thanks{These authors contributed equally.}

\affiliation{Department of Physics and Astronomy, University of Rochester, Rochester, NY, 14627 USA}

\affiliation{University of Rochester Center for Coherence and Quantum Science, Rochester, NY, 14627, USA}

\author{Yadav Prasad Kandel}

\thanks{Current address: IBM T.J. Watson Research Center, Yorktown Heights, New York 10598, USA}

\affiliation{Department of Physics and Astronomy, University of Rochester, Rochester, NY, 14627 USA}

\author{John M. Nichol}
\email{john.nichol@rochester.edu}

\affiliation{Department of Physics and Astronomy, University of Rochester, Rochester, NY, 14627 USA}
\affiliation{University of Rochester Center for Coherence and Quantum Science, Rochester, NY, 14627, USA}

\begin{abstract}
Hybrid quantum devices using surface acoustic waves show promise as key elements of quantum information processors. We report measurements of integrated flip-chip devices consisting of semiconductor quantum dots and surface acoustic wave resonators in lithium niobate. We observed that the pyroelectric effect in lithium niobate inhibited the operation of quantum dots in the integrated devices. GaAs/AlGaAs devices suffered from unintentional carrier depletion, and Si/SiGe devices suffered from electrostatic discharge. Our results highlight the importance of mitigating pyroelectric effects in semiconductor-lithium niobate hybrid devices for continued progress in quantum interconnects and transducers. 
\end{abstract}


\maketitle
\section{Introduction}
Surface acoustic wave (SAW) resonators are emerging as useful elements in modern quantum technologies, both as hosts of quantum states and interconnects~\cite{satzinger2018quantum,shao2022electrical}. 
Quantum state creation and interference effects have been demonstrated in coupled superconducting qubit-SAW systems~\cite{satzinger2018quantum, bienfait2019phonon, qiao2023splitting}. The interaction of confined surface acoustic waves has been explored in defect-spin qubits~\cite{golter2016optomechanical, golter2016coupling, whiteley2019spin}, and coherent charge~\cite{hermelin2011electrons, mcneil2011demand} and spin~\cite{bertrand2016fast} shuttling via SAW transport have been achieved in gate defined quantum dots (QDs).  Despite these successes, coherent coupling of QD charge or spin-qubits and surface acoustic phonons has not been demonstrated.

Some semiconductor substrates, such as GaAs, are natively piezoelectric, and feature the build up of charge in response to mechanical stress. Such materials would potentially enable on-chip integration of QD and SAW devices. However, these materials generally have a weak piezoelectric coupling, making it challenging to achieve the strong-coupling regime. One approach to circumvent this challenge is to combine different materials in a flip-chip architecture, which has been reported for other qubit platforms~\cite{satzinger2018quantum,satzinger2019simple}.  

Toward this end, we studied the performance of flip-chip devices designed to capacitively couple GaAs or Si QD charge or spin qubits with SAW resonators on  LiNbO$_3$ (LN). For both cases, we found that the devices were inoperable at low temperatures, likely due to the strong pyroelectricity of LN. (Pyroelectric materials feature a change in polarization as their temperature changes.) In particular, we found that electronic transport in the GaAs/AlGaAs-LN devices was depleted, likely because of a combination of the LN pyroelectricity and GaAs/AlGaAs piezoelectricity. While electronic transport in Si/SiGe-LN devices was not affected, we observed frequent electrostatic discharge effects on both the Si/SiGe and LN substrates, which we believe resulted from the LN pyroelectricity.  In total, our results suggest that either a non-capacitive coupling mechanism or a non-pyroelectric substrate should be used.  

\begin{figure}[b]
\begin{center}
\includegraphics[width=0.4\textwidth]{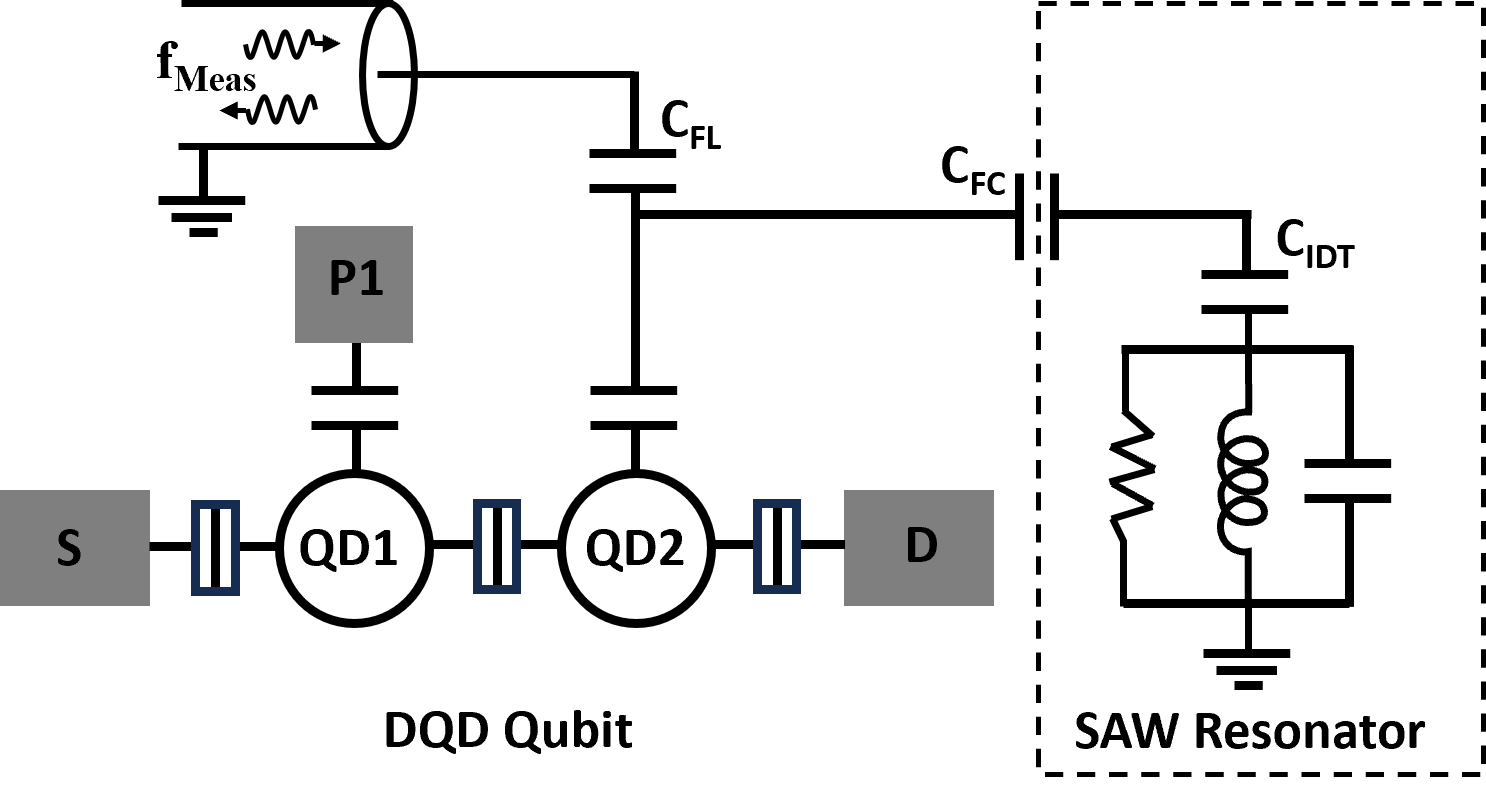}
 \caption{ Electrical schematic of a capacitively coupled DQD - SAW flip-chip device, corresponding to the QD devices fabricated in this work. The capacitances determining the effective coupling are the feed-line capacitance to the gate electrode C$_{FL}$, the inter substrate flip-chip capacitance C$_{FC}$ and the IDT capacitance C$_{IDT}$.  The DQD system is realized either in GaAs/AlGaAs or Si/SiGe, whereas the SAW resonator is fabricated on a LN substrate.}
\label{fig1}
\end{center}
\end{figure}
\section{Experimental Setup}
Capacitive coupling is the natural choice to couple QDs with microwave resonators. In such an approach, the voltage antinode of the resonator is typically galvanically or capacitively coupled to a gate electrode, which can modulate the electrochemical potential of the QD. The schematic of such a setup is shown in Fig.~\ref{fig1}.

\begin{figure}[t]
\begin{center}
\includegraphics[width=0.5\textwidth]{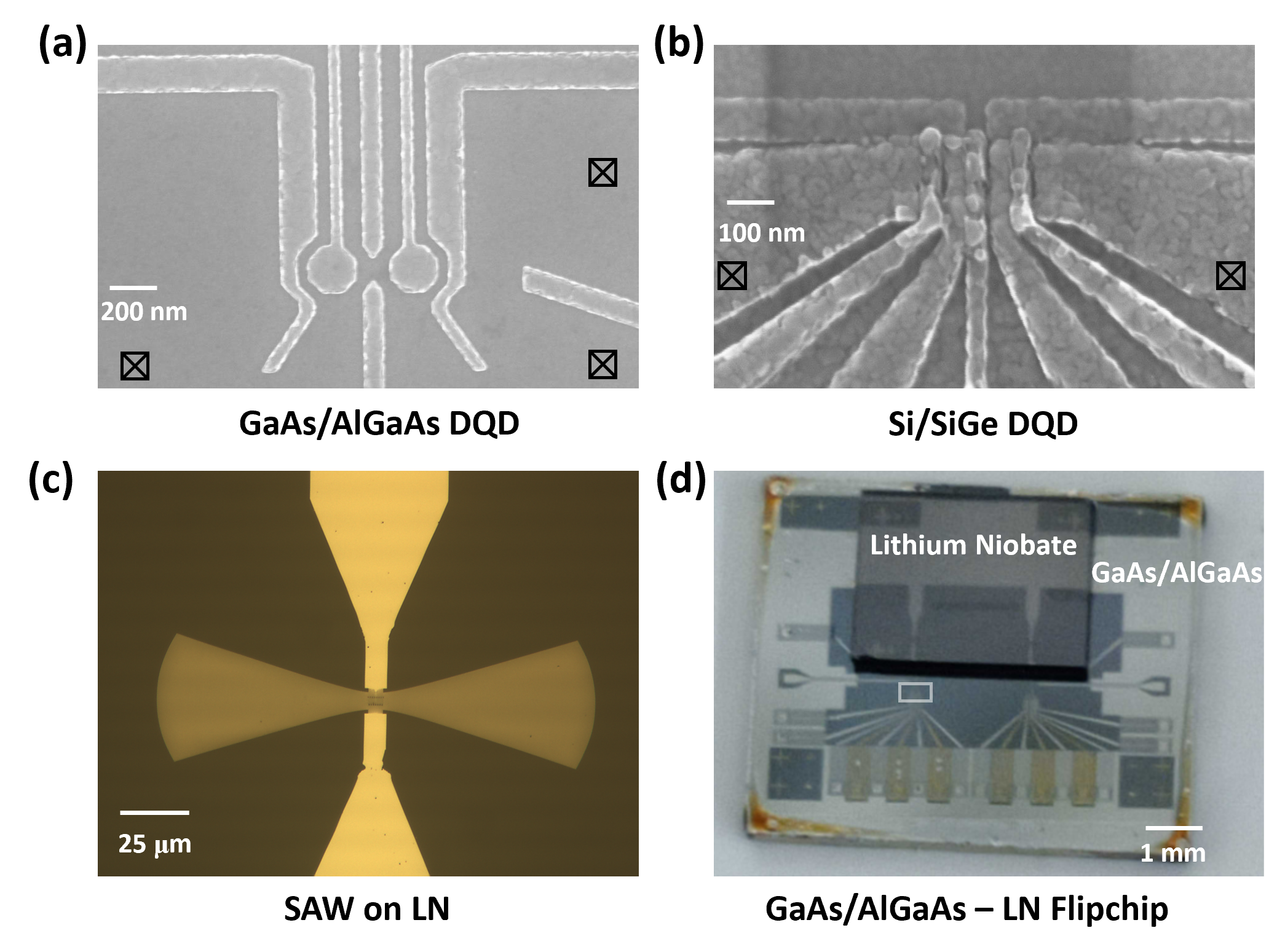}
 \caption{ (\textbf{a}) SEM image of a depletion mode DQD fabricated on a GaAs/AlGaAs heterostructure in a stadium-style geometry. (\textbf{b}) SEM image of an accumulation mode DQD fabricated on a Si/SiGe heterostructure in an overlapping-gate geometry. (\textbf{c}) Optical image of a SAW resonator fabricated on a LN substrate. (\textbf{d}) Optical image of an assembled GaAs/AlGaAs-LN flip-chip device. The white rectangle highlights the region in which the mesa containing the DQD is located.}
\label{fig2}
\end{center}
\end{figure}

\begin{figure*}[t]
\begin{center}
\includegraphics[width=1\textwidth]{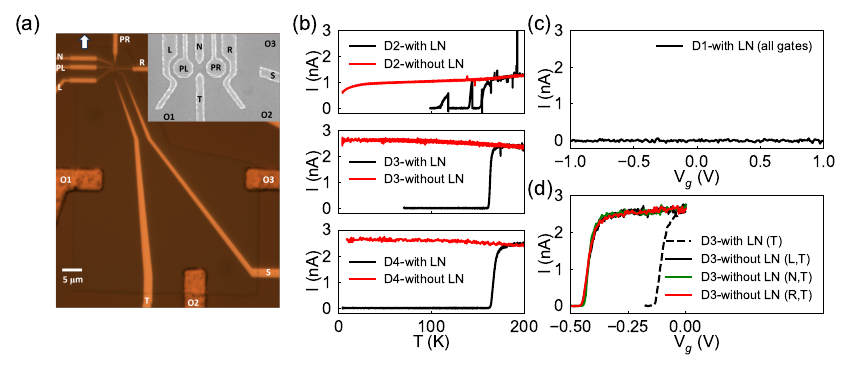}
 \caption{(\textbf{a}) Optical image of a DQD structure on a GaAs/AlGaAs substrate depicting the ohmic contacts. In an integrated flip-chip device, the LN chip would be above this image, as indicated by the white arrow. The inset shows an SEM image of the small gate electrodes. (\textbf{b}) Temperature dependence of the current between two ohmics is shown for devices D2, D3, and D4. Black curves are the response of the full flip-chip device, and the red curves are for the devices without the LN chip. (\textbf{c}) Current between ohmic contacts O1 and O2 of D1 plotted as a function of the voltage applied to all the gates shown in (a).  (\textbf{d}) Current between ohmic contacts O1 and O2 of device D3 plotted as a function of the voltage applied to the gate electrodes mentioned in the legends.}
\label{fig3}
\end{center}
\end{figure*}

Double quantum dots (DQDs) in a GaAs/AlGaAs heterostructure were fabricated in a stadium-style geometry~\cite{van2018microwave,scarlino2019all} [Fig.~\ref{fig2}(a)] using a standard process. The quantum well lies 57 nm below the surface in the heterostructures used in this experiment. One of the plunger gates terminates in a pad that capacitively couples to the SAW chip. Silicon DQDs [Fig.~\ref{fig2}(b)] were fabricated on a Si/SiGe heterostructure with an 8-nm-thick quantum well 50 nm below the surface. One side of the split screening gate~\cite{borjans2020split} terminates in a capacitor pad for coupling to the SAW chip. In the context of the results reported here, there are two major differences between GaAs/AlGaAs QDs and Si/SiGe QDs. First, GaAs/AlGaAs heterostructres are doped, and the QDs operate in depletion mode, while Si/SiGe QDs operate in accumulation mode. This means, for example, that current can flow between ohmic contacts in a GaAs/AlGaAs heterostructure without any metal gates or applied voltages. In contrast, current between ohmic contacts in Si/SiGe devices requires a gate and a positive voltage. The second major difference is that GaAs/AlGaAs is a piezoelectric substrate, while Si/SiGe is not piezoelectric. 

Surface acoustic wave resonators with resonance frequencies around 5 GHz were fabricated on  dual-side-polished black 128\degree Y-cut LN substrates.  Mirrors, interdigital transducers (IDTs), and all other structures were made of 40-nm-thick aluminum. The important features of LN for this work are its strong piezoelectricity and  pyroelectricity. Piezoelectric materials feature a change in polarization with mechanical stress, and pyroelectric materials feature a change in polarization with temperature. The key feature of the SAW resonators used here is that they are designed for high impedance~\cite{kandel2024high}, meaning that they have a small acoustic mode volume and minimal stray capacitance to other parts of the device. 

We used a technique similar to that reported by Satzinger et al.~\cite{satzinger2019simple} to integrate the QD and SAW chips in a flip-chip assembly. We first made resist ``spacers'' to set the distance between chips during assembly. We have used both hard-baked photoresist, as well as cross-linked PMMA, for this purpose. The advantage of using PMMA is that it avoids photoresist developers, which can etch aluminum and aluminum oxides.  We then used a small amount of SU-8 photoresist on the bottom substrate as glue. After that, the LN chip was aligned and compressed using a home-built mask-aligner setup~\cite{pham2018compact}. We cured the glue by baking the integrated flip-chip assembly for 2 minutes at 200\degree C. The assembled devices are rigid initially but become very fragile after a thermal cycle. The assemblies become more rigid after a few days at room temperature, enabling us to use the same device for several cooldowns. All measurements are conducted in a 4K cryostat with exchange gas.

\section{Observations}

\subsection{GaAs/AlGaAs-LN devices}
We report measurements on four integrated DQD-SAW devices, D1, D2, D3 and D4. Devices D1 and D2 were on the same flip-chip assembly, while D3 and D4 were on another flip-chip assembly. The current between two ohmic contacts was measured during cooling using a lock-in amplifier and an alternating source drain bias of 40~$\mu$V. While cooling down the samples to 4 K, we surprisingly observed that the current sharply dropped to zero at $\sim$150 K. Such sharp drops in current were observed in all the devices as shown for D2-4 in Fig.~\ref{fig3}(b) (black curves). In this ``depleted'' state at low temperature, applying positive voltages to the gates did not restore the current as shown in Fig.~\ref{fig3}(c). 

However, the current recovered after warming the devices. On repeating the thermal cycle, we observed that although most of the time the ``mesa'' (the region around the DQD) was fully depleted, there were some instances when the current did not drop to zero. However, in such scenarios, we observed that we could pinch off the current by applying a voltage to only one gate. (Normally for depletion-mode devices, voltages applied to two gates are required to deplete the current.) This observation suggest that even when current can flow, the mesa was partially depleted. 

To confirm that this effect did not relate to the specific metal features on the chips, we repeated the experiment after replacing the LN SAW chip with a bare LN chip and found the same result. Even for a GaAs device with just the mesa and ohmic contacts, we observed depletion of the current between the ohmics when LN was present.  We also found that the depletion temperature depended only weakly on the cooling rate. Compared to the standard cooldown time of $\sim$ 90 minutes, when we cooled down over a period of 5 hours, the onset temperature changed by about $\sim$5 K.  We also confirmed that this effect did not destroy the devices. After removing the LN chip, the devices exhibited the expected QD behavior [red curves in Fig.~\ref{fig3}(b)].  

In total, these results suggest that the presence of the LN chip depleted the 2DEG in the GaAs/AlGaAs heterostructure below a temperature of about 150~K. In turn, this temperature-dependent electrical behavior suggests an effect related to pyroelectricity in LN.

\begin{figure}[t]
\begin{center}
\includegraphics[width=0.5\textwidth]{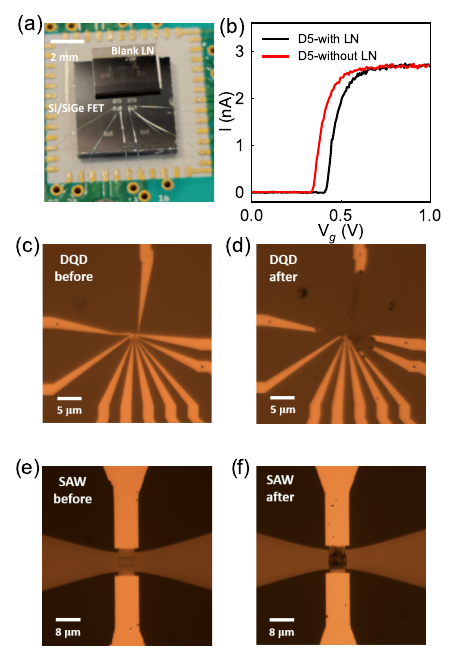}
 \caption{(\textbf{a}) Optical image of Si/SiGe-LN device D5 in which a blank LN chip is glued near a field effect transistor. (\textbf{b})  The source-drain current measured at 4 K is plotted as a function of the gate voltage for D5 with and without LN. (\textbf{c, d}) Optical images of the gates on Si/SiGe device D6 before and after cool down, respectively. (\textbf{e, f}) Optical image of the IDTs on the LN substrate of device D6 before and after cooldown, respectively.}
\label{fig4}
\end{center}
\end{figure}

\begin{figure*}[t]
\begin{center}
\includegraphics[width=0.95\textwidth]{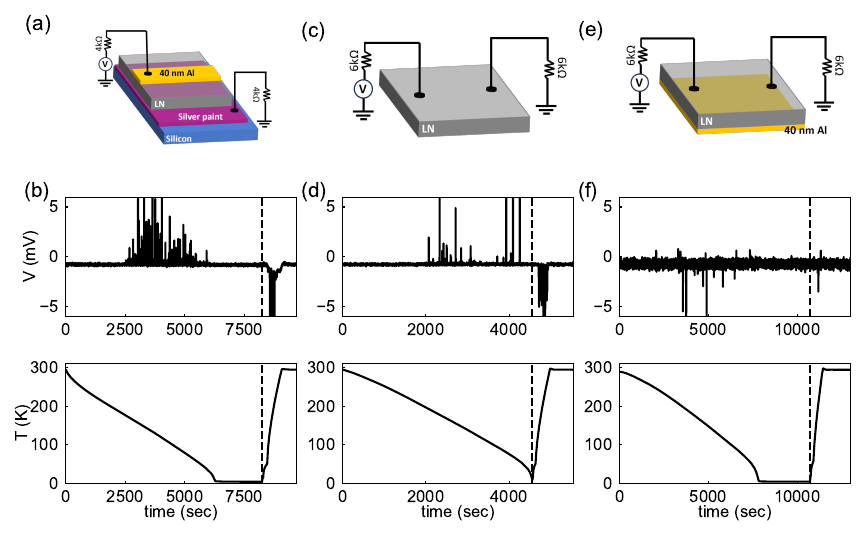}
 \caption{ (\textbf{a}) Schematic of the measurement circuit used to measure voltage differences between the two LN planes. (\textbf{b}) Measured voltage and temperature are shown as a function of time while cooling down and warming up. (\textbf{c-d}) Schematic of the measurement circuit used to measure in-plane potential differences and corresponding measured voltage and temperature while cooling down and warming up. (\textbf{e-f}) Schematic of the measurement circuit for one-side aluminum-coated LN and corresponding measured voltage and temperature while cooling down and warming up. The vertical dashed line in (b,d,f) marks the beginning of warm-up.}
\label{fig5}
\end{center}
\end{figure*}

\subsection{Si/SiGe - LN devices}
We also tested Si/SiGe-LN devices D5 and D6, hoping that these would be less susceptible to pyroelectric effects, since Si is not piezoelectric. First, we tested a field-effect transistor with and without a bare LN chip placed on top and observed a very similar turn-on response as shown in Fig.~\ref{fig4}(b). We then fabricated Si/SiGe-LN QD-SAW flip-chip devices. Surprisingly, upon cooling down these QD-SAW samples, we also could not observe turn-on behavior in the QDs, even after applying a large positive voltage. After taking out the sample from the cryostat, we observed that the small QD gate electrodes were destroyed, likely by electrostatic discharge (ESD). Interestingly, the IDT on the LN SAW device was also destroyed. However, the gates and IDTs of the other unconnected QD/SAW devices in the same assembly, which had all electrodes floating during cool down,  were not destroyed. We observed similar discharge events in multiple Si/SiGe-LN flip-chip devices. Optical images of the destroyed gates and IDT of device D6 are shown in Fig.~\ref{fig4}. 

To confirm that this effect resulted from the capacitive coupling between the QD gates and the LN chip, when we cooled down a device with the screening gate coupled to the SAW resonator grounded, the small QD gates survived the cooldown, whereas the SAW IDTs were still destroyed. These observations again hint at pyroelectricity-related effects, although their manifestation is different from the GaAs devices. 

\subsection{LN}
We performed several additional experiments on LN chips to elucidate the nature of these pyroelectric effects. We first cooled down a LN chip glued onto a Si chip with silver paint. The top side of the LN chip was partially coated with a 40-nm-thick Al film as shown in Fig.~\ref{fig5}(a). In this way, we could measure the voltage difference between the two planes of the LN crystal. During cooling, we started observing voltage spikes below 200 K that disappeared at base temperature ($\sim$4 K) as shown in Fig.~\ref{fig5}(b). Such spikes were previously observed by other groups in LN and were attributed to pyroelectricity \cite{thiele2024pyroelectric}. While warming up, the voltage spikes appeared with the opposite polarity and vanished once the sample reached room temperature [Fig.~\ref{fig5}(b)]. Next, we cooled down a blank LN chip to measure any on-plane voltage differences, as shown in Fig.~\ref{fig5}(c). In this case also, similar voltage spikes were observed during both cool down and warm up [Fig.~\ref{fig5}(d)]. 

The pyroelectric effect is known to be suppressed when the entire surface is coated with a conducting film. We tested a blank LN chip coated with a 40-nm-thick aluminum film on the bottom surface [Fig.~\ref{fig5} (e)] and observed that the number of voltage spikes was reduced, as shown in Fig.~\ref{fig5} (f), and compared with Fig.~\ref{fig5} (d). (Note that Figs.~\ref{fig5}(d) and (f) compare intra-plane voltage differences, while Fig.~\ref{fig5}(b) displays inter-plane voltage differences.) While our observations indicate a reduction in intra-plane voltage spikes when the the opposite plane is fully  coated with metal, further experiments are required to reach a definite conclusion about the effect of metalization on the intra-plane voltage differences. 

\section{Discussion}
We suggest that our observations can be explained by the pyroelectricity of LN together with the material properties of the GaAs/AlGaAs and Si/SiGe devices. In particular, we suggest that the pyroelectric polarization of the LN generates a large electric field, which causes strain in the GaAs, which in turn depletes the 2DEG. For the Si/SiGe devices, we suggest that the large electric fields developed through the pyroelectric polarization can lead to dielectric breakdown and electrostatic discharge. 

Strictly speaking, a pyroelectric material changes its polarization in response to a change in temperature. However, this change in polarization has recently been used to  gate graphene devices, suggesting that in some instances, this polarization can generate a voltage difference or electric field~\cite{mou2025unexpected}. Indeed, in our measurements we measured a pyroelectric voltage difference between the two surfaces of a LN chip [Fig.~\ref{fig5}(b)]. The polarization or voltage may also vary spatially, leading to voltage gradients and current spikes on the same crystal plane.  Similar spikes in current measured between co-planar electrodes on LN were observed by Thiele et. al.~\cite{thiele2024pyroelectric}, and those current spikes were later correlated with changes in optical properties of LN~\cite{johnston1970optical,schein1978pyroelectric,kostritskii2012optical}.

Assuming that the pyroelectric polarization generates an electric field, we hypothesize that this field creates a strain field in the GaAs devices mediated by its intrinsic piezoelectricity. We also suggest that this strain depletes the two-dimensional electron gas.  The proposed process is schematically shown in Fig.~\ref{fig6}. This hypothesis is consistent with previous reports, which show that stress can deplete GaAs/AlGaAs electron gasses~\cite{fung1997linear,fung1998plane,liu2001uniaxial}.

It is surprising, however, that in our devices, the depleted region is not directly beneath the LN, but rather hundreds of micrometers away from its edge. To explain this, we hypothesize that the strain relaxes over a length scale corresponding to the lateral size of the LN chip. Indeed, we have observed that the GaAs/AlGaAs mesa appears ``less depleted'' at longer distances from the LN chip. For example, when we occasionally observed non-zero current at cryogenic temperatures in GaAs/AlGaAs-LN devices, we also observed that the ``T'' and ``S'' gates by themselves could pinch off the current [see Fig.~\ref{fig3} (a)]. If we suppose that only the upper part of the mesa is depleted, only these two gates would be able to pinch off the current. 

To explain our observations from the Si/SiGe-LN devices, we again hypothesize the existence of a large potential difference between the SAW device and the QD device as a result of the pyroelectric polarization. Because of this large potential difference, the segment of the screening gate that is capacitively coupled to the SAW IDT can rise to a large potential with respect to the grounded gates, triggering dielectric breakdown on the QD chip, and destroying the small gate electrodes in the process. When this initial breakdown occurs, the magnitude of the voltage difference across the SAW IDT increases, triggering a second breakdown event in the SAW device. To partially test this hypothesis, we verified that grounding the QD gate coupled to the SAW prevented breakdown on the QD device but still allowed breakdown of the SAW IDT. 

Surprisingly, the breakdown effects were never observed in GaAs/LN devices. A possible explanation for this is partial screening of the pyroelectric field from the piezoelectric polarization in GaAs. It could also be related to a secondary pyroelectric effect~\cite{schein1978electrostatic,nix1988primary,chew2003primary,liu2025bond}, which depends on the difference in thermal expansion between the two substrates. The tertiary pyroelectric effect~\cite{schein1978electrostatic2,kosorotov1986tertiary}, which is related to the temperature gradient within the LN substrate, might also play a role. 

We emphasize that our data can neither confirm nor refute the hypothesis we have presented but instead serve as the starting point for future work. Alternative hypotheses for our results could involve a mismatch between the coefficients of thermal expansion between LN and the different semiconductors, together with large piezoelectric voltages, or perhaps long-range electric fields from the build up of charge in the LN.  

\section{Future Directions}
Future work to confirm our hypotheses could involve nanofabricated strain gauges on the GaAs/AlGaAs-LN devices to verify the proposed mechanism involving depletion via strain. Detailed finite element simulations of the combined system could potentially be used to elucidate the mechanism of depletion. Voltage measurements on the Si/SiGe-LN devices could also confirm the building up of large voltage gradient across the QD gates and the SAW IDT during cooldown.

Assuming that pyroelectricity in LN is the main cause for our observations, reducing the pyroelectric polarization or its effects is critical for integrated LN-semiconductor devices of the type we have described here. It could be that different crystal orientations have a lower polarization. Another strategy is to deposit metal features on the LN to hold the surface at a fixed potential~\cite{thiele2024pyroelectric}. However, the introduction of additional metalization will increase the stray capacitance, potentially necessitating a shift to inductive QD-SAW coupling mechanisms, as has previously been explored in the context of superconducting qubits~\cite{satzinger2018quantum}.  It is also worthwhile to consider how the pyroelectric polarization could be harnessed in potential pyroelectric-piezoelectric hybrids~\cite{yang2020piezoelectric,gaur2024exploiting} or through voltage-free gating.~\cite{mou2025unexpected}.

\begin{figure}[t]
\begin{center}
\includegraphics[width=0.5\textwidth]{Figure6.png}
 \caption{ Schematic of the possible interplay of pyroelectricity in LN and piezoelectricity in a GaAs/AlGaAs heterostructure (left: cross-section view, right: top view). Temperature changes generate charged surfaces due to pyroelectricity. These charged surfaces generate an effective electric field, which  results in strain in the GaAs heterostructure. The resulting stress, because of GaAs piezoelectricity, can deplete the 2DEG in the GaAs heterostructure over a relatively large distance. The double arrows indicate stress, and their length is a qualitative indicator of the stress magnitude.}
\label{fig6}
\end{center}
\end{figure}

\section{Conclusion}
In summary, we have reported measurements of hybrid QD-SAW devices in GaAs/AlGaAs-LN and Si/SiGe-LN flip-chip architectures. In both cases, we could not operate QDs as expected, likely because of effects related to pyroelectricity in LN. In particular, we hypothesize that the pyroelectric polarization in LN depleted carriers in GaAs/AlGaAs devices and destroyed QD gates through ESD in Si/SiGe devices. Going forward, mitigating effects related to pyroelectricity in semiconductor-LN structures will be critical for exploring charge- and spin-phonon coupling in quantum-dot-based quantum systems. 

\section{ACKNOWLEDGMENTS}
MRS acknowledges several discussions with Habitamu Walelign and Jake Markowski during the fabrication of the QD devices. This work was sponsored by the Office of Naval Research Office through Grant No. N00014-20-1-2424 and the Army Research Office through Grant No. W911NF-19-1-0167. The views and conclusions contained in this document are those of the authors and should not be interpreted as representing the official policies, either expressed or implied, of the Army Research Office or the U.S. Government. The U.S. Government is authorized to reproduce and distribute reprints for Government purposes notwithstanding any copyright notation herein.

\section{Appendix A: Fabrication details}

DQDs in GaAs/AlGaAs heterostructure were fabricated in a stadium style geometry. The mesa region was created by chemical etching using a 4:4:100 ml mixture of H$_2$O$_2$:H$_3$PO$_4$:H$_2$O. The ohmic contacts were made out of Ni/Ge/Au/Ni/Au (6/30/60/20/150 nm). The samples were annealed at 480 \textdegree C for 6 minutes to achieve ohmic contact with the 2DEG. 10 nm of Al$_2$O$_3$ was deposited using atomic layer deposition before making the gate electrodes. The inner parts of the gate electrodes were made of 50 nm aluminum. The outer parts of the gate electrodes, including the bonding pads, were made in another layer with 200 nm aluminum, which was deposited after argon ion milling for 30 seconds to remove surface oxides. LC filters close to the bond pads [Fig.~\ref{fig2} (d)] were made in the initial batch of devices to reduce microwave losses, as discussed extensively by Mi et al. in Ref.~\cite{mi2017circuit}. These filters were not made in the devices described in this paper.  The first batch of devices fabricated without the Al$_2$O$_3$ insulating layer showed high noise and high hysteresis with gate voltages, as reported in an earlier work~\cite{liang2020reduction}.

DQDs on the Si/SiGe heterostructure were fabricated with an overlapping gate geometry. The mesa region was created by plasma etching (SF$_6$+O$_2$). The ohmic contacts were realized by phosphorous ion implantation followed by annealing at 700 \textdegree C for 5 minutes and depositing 5/70 nm of Cr/Au. A 15-nm-thick Al$_2$O$_3$ insulating layer was deposited before making the aluminum gate electrodes. The gate electrodes were made in 3 steps: the screening gates, followed by the accumulation gates and the plunger gates, and finally the barrier gates with thicknesses of 30/40/50 nm, respectively, deposited via thermal evarpoation. The first 3 nm of aluminum in each gate layer is deposited at a very low rate of $\sim$0.1 \AA/s, and the remaining is deposited at a rate of $\sim$1 \AA/s. The top screening gate was designed to have two halves, and one side was used to capacitively couple the DQD to the SAW resonator\cite{borjans2020split}.

\section{Appendix B: Characterization of SAW resonators in the flip-chip devices}

\begin{figure}[t]
\begin{center}
\includegraphics[width=0.48\textwidth]{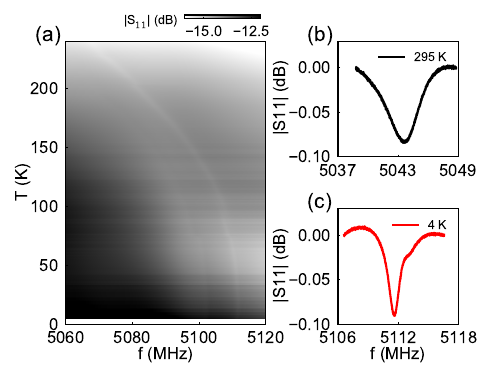}
 \caption{(\textbf{a}) |$S_{11}$| plotted as a function of frequency and temperature showing the increasing SAW resonance frequency with lowering temperature. (\textbf{b}) The SAW resonance response (background corrected) at room temperature. (\textbf{c}) The SAW resonance response (background corrected) at 4K showing the enhanced quality factor. All the data shown were measured in the device D1.}
\label{fig7}
\end{center}
\end{figure}

We characterized the SAW resonators in the GaAs/AlGaAs - LN flip-chip devices as a function of temperature. Fig.~\ref{fig7} (a) shows the temperature dependence of the SAW resonance frequency measured in the flip-chip device D1. The resonance at room temperature appears very close to the designed value, and it shifts to higher frequency when the temperature is lowered. This increase in frequency has been previously reported~\cite{yamamoto2023low, magnusson2015surface, li2015aln, el1996high, tarumi2012low, emser2022minimally, li2018temperature, manenti2016surface, chen2020ultrahigh} and is attributed to the combined effect of thermal contraction and change in the dielectric constant of the substrate. The observation of the  SAW resonance in the integrated flip-chip device, where the scattering parameter $S_{11}$ is measured through a CPW line on the GaAs/AlGaAs wafer, demonstrates proper operation of the inter-chip capacitive coupling. We did not observe any unwanted jumps in the resonance frequency, probably indicating that the pyroelectric charge accumulation does not significantly impact the surface acoustic properties.

\bibliography{bib}

\end{document}